\documentclass[conference]{IEEEtran}
\IEEEoverridecommandlockouts

\usepackage{cite}
\usepackage{amsmath,amssymb,amsfonts}
\usepackage{algorithmic}
\usepackage{graphicx}
\usepackage{textcomp}
\usepackage{xcolor}
\usepackage{balance}
\usepackage{comment}
\usepackage[ruled,linesnumbered]{algorithm2e}
\SetKwProg{Init}{Initialization:}{}{}

\usepackage{url}

\usepackage{enumitem}

\begin{document}

\title{ADR-Lite: A Low-Complexity Adaptive Data Rate Scheme for LoRa Network}

\author{Reza Serati$^*$, Benyamin Teymuri$^*$, Nikolaos Athanasios Anagnostopoulos$^\dagger$, and Mehdi Rasti$^{*, **}$ 
\\ $^*$Department of Computer Engineering, Amirkabir University of Technology, Tehran, Iran
\\ $^\dagger$Faculty of Computer Science and Mathematics, University of Passau, Passau, Germany 
\\ $^{**}$Centre for Wireless Communications, University of Oulu, Finland
\\ $^*$Emails: {\{re.serati, benyamin.teymuri, rasti\}@aut.ac.ir}
\\$^\dagger$Email: Nikolaos.Anagnostopoulos@uni-passau.de
$^{**}$Email: mehdi.rasti@oulu.fi
}
\maketitle

\begin{abstract}
The Internet of Things (IoT) is currently used for various applications, including smart cities, agriculture, and smart homes. The IoT applications’ long-range and low energy consumption requirements have led to a new wireless communication technology known as Low Power Wide Area Network (LPWANs). In recent years, the Long Range (LoRa) protocol has gained a lot of attention as one of the most promising technologies in LPWAN. Choosing the right combination of transmission parameters is a major challenge facing the LoRa network. LoRa executes an Adaptive Data Rate (ADR) mechanism to configure each End Device’s (ED) transmission parameters, resulting in improved performance metrics. In this paper, we propose a linkbased ADR approach that aims to configure the transmission parameters of EDs by making a decision without taking into account the history of the last received packets, resulting in a relatively low space complexity approach. In this study, we present four different scenarios for assessing performance, including a scenario where mobile EDs are considered. Our simulation results show that in a mobile scenario with high channel noise, our proposed algorithm’s Packet Delivery Ratio (PDR) is 2.8 times outperforming the original ADR and 1.35 times that of other relevant algorithms.
\end{abstract}
\section{Introduction}
A consistent low-cost and low-energy connectivity amongst all smart devices is required to build an intelligent society~\cite{iot}. In the Internet of Things (IoT) environment, Low Power Wide Area Networks (LPWANs) are developed for energy consumption optimization and improved communications range. The Packet Delivery Ratio (PDR), Energy Consumption (EC), resilience in the face of faults and challenges, and coverage area are some measures that may be used to assess a network's performance. Environmental conditions such as urban (UR) and suburban (SU) conditions, the number of transmitting end devices (EDs), the number and placement of Gateways (GWs), network topology, and regulatory restrictions are salient factors that can directly influence network functionality~\cite{lora-study}. The LoRa network is a low-power, long-range communication protocol that can cover a wide distance. To establish a communication link, a set of transmission parameters have to be configured. Transmission parameters such as the Spreading Factor (SF), Transmission Power (TP), Carrier Frequency (CF), Bandwidth (BW), and Coding Rate (CR) can be configured in a LoRa network to ensure reliable communication.
\par Combining the transmission parameters provides a state space from which hundreds of configurations can be chosen, impacting the network performance ~\cite{lora-probing}. Choosing the right combination of transmission parameters is a major challenge facing the LoRa network. In the central decision-making Network Server (NS), LoRa executes an Adaptive Data Rate (ADR) mechanism to configure EDs' transmission parameters, resulting in improved performance metrics. 
To increase the efficiency and scalability of LoRa networks, numerous articles with different approaches have been published. Changes in Media Access Control (MAC)~\cite{lora-cad}, the number of message retransmissions~\cite{msg-repli}, statistical and mathematical models~\cite{lora-queue},%
optimization algorithms~\cite{adr-owa}, and machine learning techniques~\cite{adr-mixmab} are among the approaches discussed.
This paper aims to review, implement, and analyze a new greedy approach while maintaining minimal space complexity. This approach can improve network performance in terms of reducing the collision rate and thereby increasing PDR. Contributions made by this work are as follows:
\begin{itemize}[after=\vspace*{-1pt}]
 \item Our proposed \textbf{A}daptive \textbf{D}ata \textbf{R}ate \textbf{L}ow-complex\textbf{i}\textbf{t}y sch\textbf{e}me, ADR-Lite, configures the transmission parameters of the LoRa network in variable channel conditions, independent of the EDs' number and distribution, whether they are static or on the move. This is achieved while our algorithm's space complexity remains optimized compared to other approaches.
 \item Unlike existing approaches, which are limited to  setting SF and TP only, our suggested algorithm includes adjusting SF, TP and  other transmission parameters such as CF and CR. Thus, ADR-Lite offers a greater set of configuration parameters than other ADR schemes, making it more flexible and adaptable to various deployments and requirements.
\item Our simulation results show that our proposed ADR-Lite improve the ratio of total consumed energy by all EDs to the PDR, in different scenarios.\\
\end{itemize}
\par The rest of this paper is structured as follows. The background and related works are presented in Section II. Section III describes our suggested greedy algorithm. The simulation results and conclusion are included in sections IV and V, respectively.
\section{Background and Related Works}
This section reviews the LoRa physical layer and transmission parameters. Then, it examines the related works. 
\subsection{LoRa overview}
\begin{figure}[t]
    \centering
    \includegraphics[width=\linewidth]{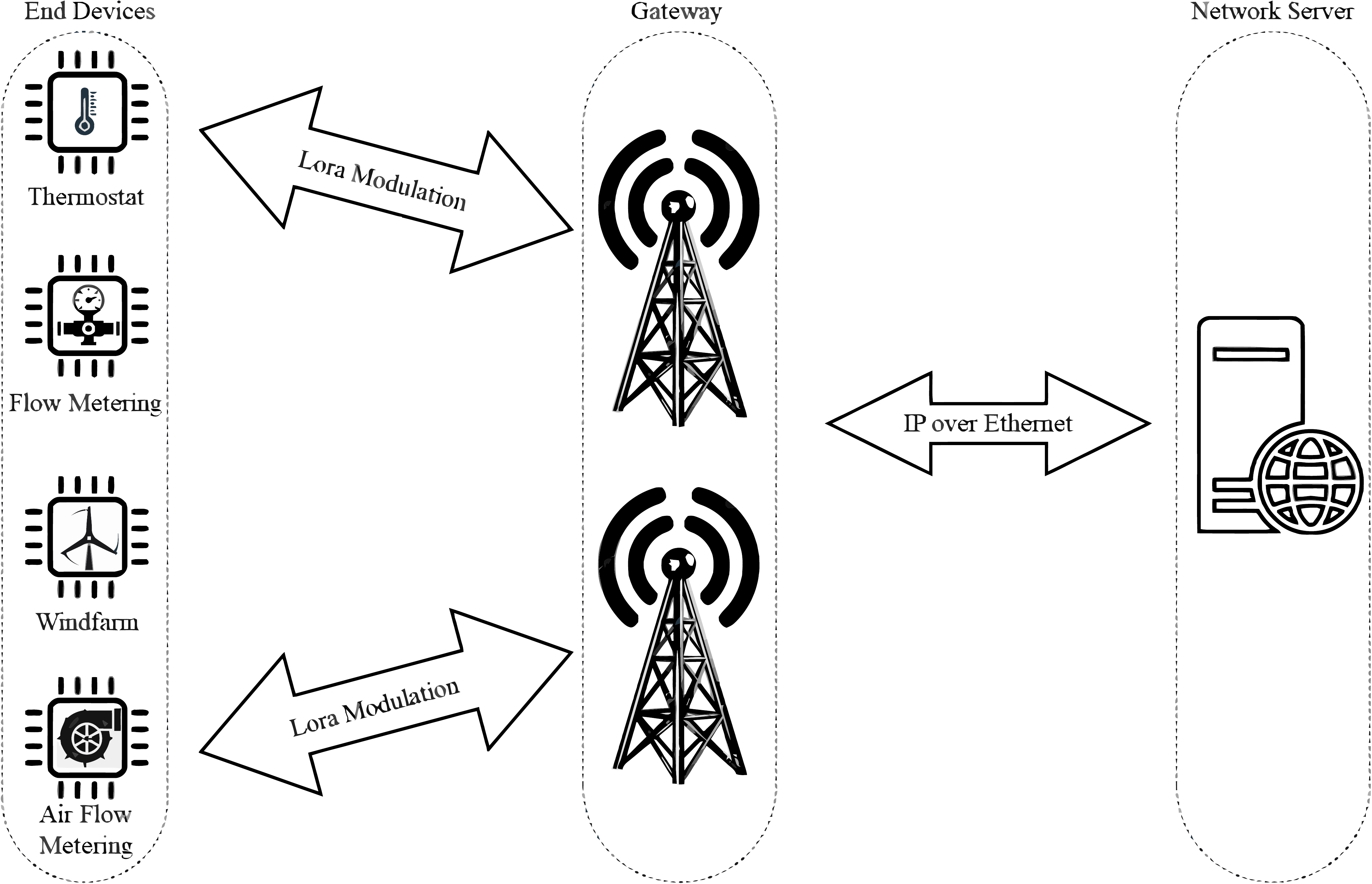}
    \caption{LoRaWAN network architecture.}
    \label{fig:arc}
\end{figure}

The LoRa protocol is a proprietary technology for a long-range, low-power network. This method employs the Chirp Spread Spectrum (CSS), which is one of the spread spectrum modulations. LoRa wireless devices have become an important part of the wireless IoT infrastructure. It has low efficiency in terms of bits per second, while it can transmit no more than 255 bytes of data per packet, which is sufficient for many IoT applications~\cite{lora-limits}. The LoRa architecture employs the star-of-star topology, which has three types of devices seen in Figure~\ref{fig:arc}.
\par LoRa makes use of unlicensed sub-gigahertz radio frequency bands like the 433, 868, or 915 MHz industrial, scientific, and medical (ISM) bands, as determined by the region it is deployed in~\cite{lora-study}. There are five transmission parameters that can be configured for appropriate sender-receiver communication, impacting communication link quality. The descriptions of these parameters are listed below~\cite{lora-study}:
\begin{itemize}
    \item \textit {\textbf{SF:}} 
    In the spread spectrum LoRa modulation, each bit of the payload message represents multiple chips of information. The symbol rate is the rate at which the spread information is sent. SF is the ratio of the nominal symbol rate to the chip rate, which represents the number of symbols transferred per bit of data. The SF can be selected from 7 to 12. Note that since different spreading factors are orthogonal to one another, the SF must be known in advance on both the transmitting and the receiving sides of the link.
    \item \textit {\textbf{TP:}}
    The amount of power that an ED should put in to transfer its messages is known as transmission power. The LoRa radio TP may vary in steps of $3$ dBm from $2$ to $14$ dBm.
    \item \textit {\textbf{CF:}}
    The central frequency, measured in Hertz, of a carrier wave that is modulated to transmit signals, is known as the carrier frequency. CF may be configured in the frequencies of 433, 868, and 915 MHz with different step sizes depending on the LoRa chip and the regulation rules.
    \item \textit {\textbf{BW:}}
    Bandwidth is the frequency range between the lowest and highest frequencies that can be reached without causing signal power degradation. Increasing the signal bandwidth allows for a higher data rate and lower transmission time at the expense of reducing sensitivity. LoRa employs the bandwidth ranges of 125 kHz, 250 kHz, and 500 kHz.
    \item \textit {\textbf{CR:}} 
    To enhance the link's robustness even further, LoRa uses cyclic error coding for forward error detection and correction. Such error coding increases the transmission overhead, reducing the data rate and improving the link's reliability in the presence of interference. The CR, and hence the interference resistance, may be changed according to the channel conditions. The values of CR may vary in the range of $\{\frac{4}{5}, \frac{4}{6}, \frac{4}{7}, \frac{4}{8}\}$.
\end{itemize}
\par The combination of these parameters can affect the data rate, noise resistance, receiver sensitivity, packet transmission delay, and energy consumption. For example, if ED is configured to transmit with the maximum SF, i.e., SF12, the most extended coverage area, the highest energy consumption, the longest transmission delay, and the lowest data rate will be achieved~\cite{lora-probing}. Consequently, finding the best combination of transmitting parameters is one of the main challenges of optimizing LoRa networks to reduce power consumption and collision rates while increasing PDR.
\subsection{Related Works}
\par Many relevant works have attempted to improve LoRa network performance~\cite{lora-probing, adr-owa, adr-mixmab, adr-avg, adr-max-lora-specification}. Examining the full transmission parameter state space to find the best combination is an exhaustive approach. A probing approach is described in~\cite{lora-probing}, with the objective of finding an appropriate combination with the least amount of state space exploration. Along with minimizing the size of the space state, the suggested strategy also considers lowering energy usage.
\par The approaches available for controlling and managing the transmission parameters in the LoRa network are divided into two categories, network-aware~\cite{adr-mixmab} and link-based approaches~\cite{adr-owa, adr-avg}. The transmission parameters between the EDs and the GWs are determined by the NS in a centralized manner (link-based approach), whereas the transmission parameters are determined in a distributed manner by the EDs, in the network-aware approach.
\par The ADR algorithm is a mechanism to adjust the transmission parameters of EDs that not only can reduce the ED's energy consumption but also results in better PDR. The default ADR method, which is a link-based approach known as ADR-MAX, uses the maximum value of the last 20 packets' Signal to Interference and Noise Ratio (SINR) as an indicator to evaluate the quality of the link~\cite{adr-max-lora-specification}. ADR-MAX makes optimistic decisions and sends an acknowledgment to the ED, regardless of the circumstantial network parameters, i.e., the ED's densification and channel saturation.
\par The authors of~\cite{adr-avg}, which proposes the FLoRa framework, intend to change the decision-making mechanism from the maximum SINR to the average SINR by offering a solution called ADR-AVG to improve network performance. Instead of optimistic decision-making, a new approach called ADR-OWA uses the ordered weighted averaging operator in~\cite{adr-owa}, dynamically configuring SF and TP based on the channel condition. In addition to the described methods that evaluate the communication link, there's also a scheme known as No-ADR, in which all transmission parameters are determined at random, meaning no link-based or network-aware approaches are used.
\par However, there are several shortcomings to these approaches. The first disadvantage is that, despite the EDs' densification in both static and mobile scenarios, the PDR measure would be deficient if ADR-MAX was adapted. The second drawback is that employing ADR-AVG would result in poor PDR where there is a low number of the EDs in the area. Furthermore, in a mobile scenario, this method's performance would be worse compared to the case of a random selection scheme. The third shortcoming is that when the noise level in the environment is low, the PDR of ADR-OWA would be less efficient than when the average SINR of the last 20 packets is used. Lastly, owing to its randomness and to not taking into account environmental changes, the No-ADR scheme will result in the least PDR of all the approaches.
\section{Our proposed ADR-Lite algorithm} 
As pointed out in the previous section, using the maximum and average SINR of the last 20 packets to evaluate the link quality or determining transmission parameters randomly will cause poor performance in networks with varying channel conditions. Thus, we propose a link-based ADR algorithm that attempts to configure the transmission parameters of EDs without considering the previous packet history, resulting in a low-space-complexity algorithm.
\par To illustrate the energy consumption model for LoRa EDs, in the following, we utilize the energy consumption formulas presented in~\cite{lora-energy}. As described in~\cite{lora-datasheet}, the consumed energy in the data transmission mode, $E_{ToA}$, is higher than in the other modes of the LoRa energy consumption model, and is calculated as follows: 
\begin{equation} 
    E_{ToA}=\left(P_{ON}\left(f_{MCU}\right)+P_{ToA}\right) \times T_{ToA}\,,
    \label{e-toa}
\end{equation}
where $P_{ON}(f_{MCU})$ is the micro controller's energy consumption depending on its processor frequency, $f_{MCU}$, while $P_{ToA}$ and $T_{ToA}$ are the dissipated energy in the transmission mode and its time duration, respectively~\cite{lora-energy}. $T_{ToA}$ is the required time for transferring both the preamble and the payload message, i.e., $T_{preamble}$ and $T_{payload }$, respectively. $T_{preamble}$ can obtained as follows:
\begin{equation} 
    T_{\text {Preamble}}=\left(4.25+N_{P}\right) \times T_{\text {symbol}}\,,
    \label{e_pre}
\end{equation}
where $N_{P}$ is the number of preamble symbols, while $T_{payload}$ is calculated by multiplying $T_{\text {symbol}}$ and $N_{\text {Payload}}$, which can be calculated as follows:
\begin{equation} 
        T_{symbol}=\frac{2^{SF}}{BW}\,,
    \label{t_symb}
\end{equation}
and
\begin{equation} 
    N_{\text{Payload}}=8+\max\left( \Bigl\lceil{\frac{\Theta(PL, SF)}{\Gamma(SF)}} \Bigr\rceil  \times{\frac{1}{CR}} , 0 \right)\,,
\label{n_pay}
\end{equation}
respectively. Here, $\Theta(PL,SF)$ is defined by:
\begin{equation} 
    \Theta(PL,SF)=8 \times PL-4 \times SF+16+28-20 \times H\,,
\label{theta}
\end{equation}
where $PL$ is packet payload length, and $H$ is equal to zero when the header is enabled or to one when there is no header present. Also $\Gamma(SF) = SF-2 \times DE$, wherein $DE$ is set to one when the low data rate optimization is enabled; otherwise, $DE$ is set to zero.
\par We assume the LoRa network consists of $U$ EDs forming a set $\mathcal{U}=\{1,2,\ldots,U\}$. The transmission parameters SF, TP, CF, and CR are represented by a configuration array, such that $I_k=\{SF_k,TP_k,CF_k,CR_k\}$. Further, there are $|\mathcal{K}|$ configuration arrays, denoted by $\mathcal{K}=\{I_1, I_2, \ldots, I_{|\mathcal{K}|}\}$. In the NS, for each ED, there is a sorted $\mathcal{K}$ in an ascending manner based on equations (\ref{e-toa} - \ref{n_pay}). Let $k_u(t)$ be the $k$th (configuration) set of parameters that has been chosen by the NS for the $u$th ED at the $t$th iteration, and let $r_u(t)$ be the configuration of the last received packet from the $u$th ED at iteration $t$. The very first $k_u(t)$ is selected to be equal to $|\mathcal{K}|$ to maximize the probability of successful communication. Lastly, two auxiliary variables named $min_u$ and $max_u$ are used to determine the future values of $k_u(t)$ as follows.
\par
\begin{figure}[t]
    \centering
    \includegraphics[width=\linewidth]{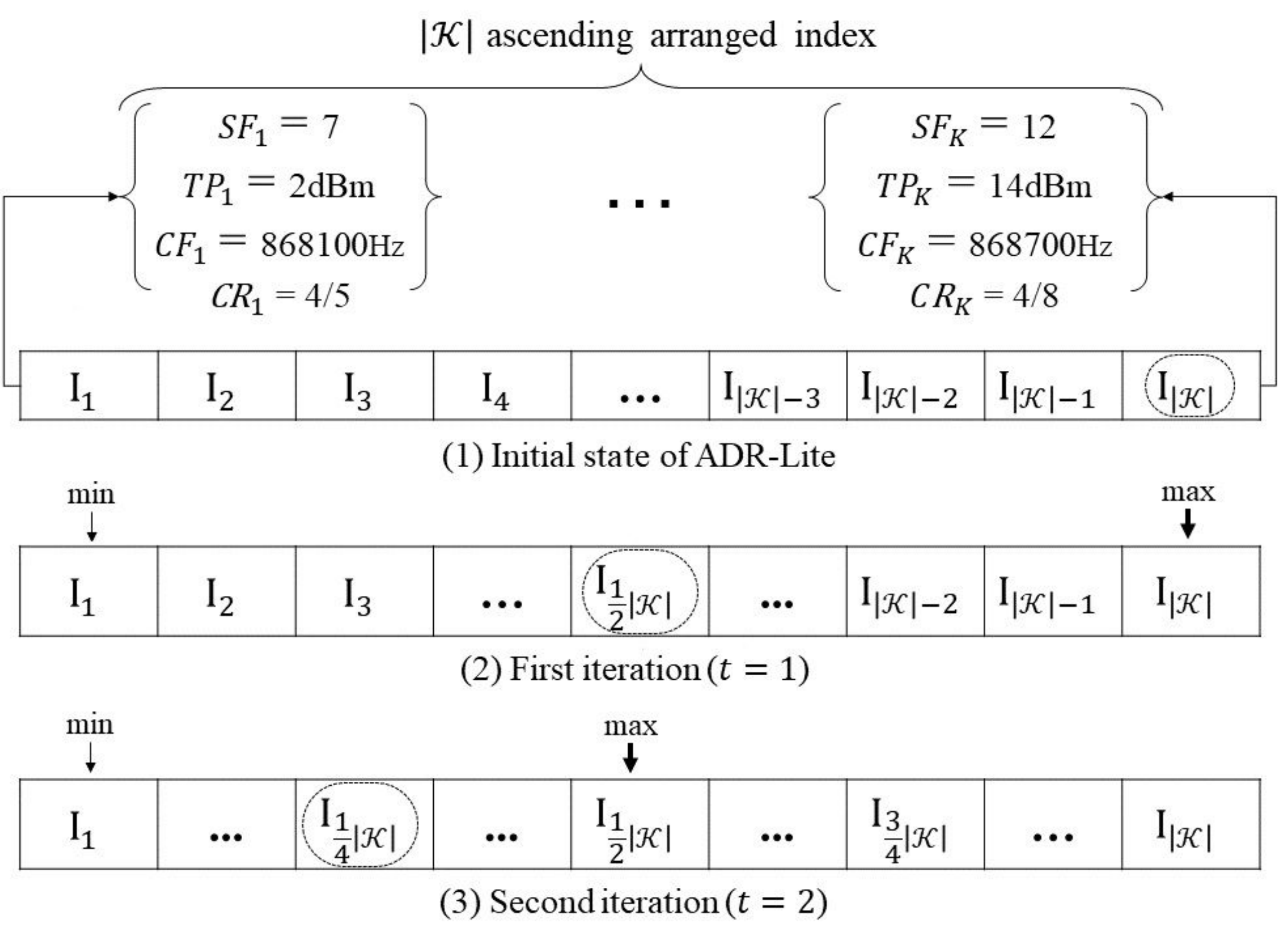}
    \caption{Possible early stages of proposed ADR-Lite scheme for the $u$th ED.}
    \label{fig:lite}
\end{figure}

 For each ED $u$, Algorithm \ref{lite} describes the proposed solution that runs on the NS during the simulation. Lines 1--8 describe the initialisation phase that is run only once for each ED, through line 10 of the Algorithm, while lines 11--20 describe the selection of  of $k_u(t)$ for iteration $t$. During each iteration $t$, the ED configures its transmission parameters based on $k_u(t)$, which has been passed on from the NS. The value of $k_u(t)$ corresponds to 
the most suitable configuration for that particular ED on that specific iteration considering the environmental conditions. In the NS, $k_u(t)$ will be determined as half of its previous value, if the configuration of the last received packet from the $u$th ED at iteration $t$, denoted by $r_u(t)$, is equal to $k_u(t-1)$, the previous value of the index. Otherwise, $k_u(t)$ will be set as the average of $k_u(t-1)$ and $|\mathcal{K}|$. Figure \ref{fig:lite} represents the possible early stages of the proposed ADR-Lite scheme for the $u$th ED.
\begin{algorithm}[t!]
    \label{lite}
    \SetAlgoLined
    \SetKwInOut{Input}{input}
    \SetKwInOut{Output}{output}
    \Input{
    $k_u(t-1)$, $r_u(t)$
    }
    \Output{
    $k_u(t)$
    }
    \Init{}{
    \vspace{2pt}
    Set $u \in \mathcal{U}$ to be the $u$th ED\\
    \vspace{2pt}
    Set $I_k=\{SF_k,TP_k,CF_k,CR_k\}$\\
    \vspace{2pt}
    Set $|\mathcal{K}|$ to be the total number of configurable parameters, so that $\mathcal{K}=\{{\textit{I}_{1},\textit{I}_{2},\ldots,\textit{I}_{|\mathcal{K}|}}\}$\\
    \vspace{2pt}
    Sort $\mathcal{K}$ ascending according to EC\\
    \vspace{2pt}
    Set $k_u(t)$ to be the index of the $I_{k_u(t)} \in \mathcal{K}$ that has been chosen for the $u$th ED at the $t$th iteration\\
    \vspace{2pt}
    Set $r_u(t)$ to be the index of the configuration $I_{r_u(t)} \in \mathcal{K}$ used in the last received packet from the $u$th ED at iteration $t$\\
    \vspace{2pt}
    Set $k_u(0) = |\mathcal{K}|$\\
    }
    \vspace{2pt}
    \For{the $uth$ ED}{
    \vspace{2pt}
    Initialization\\
    \vspace{2pt}
    \For {$t=1$ \textnormal{to} $T$
    \vspace{2pt}
    }
    {
    \vspace{2pt}
    \eIf{$r_u(t) = k_u(t-1)$}
    {
        $min_u = 1$\\
        $max_u = k_u(t-1)$
    }
    {
        $min_u = k_u(t-1)$\\
        $max_u = |\mathcal{K}|$
    }
    $k_u(t) = {I_{\left\lfloor\frac{\scriptstyle max_u+min_u}{\scriptstyle 2}\right\rfloor}}$
    \vspace{2pt} 
    }
    }
    \caption{ADR-Lite on NS}
\end{algorithm}
\section{Numerical Results}
Our algorithm's performance is evaluated by simulation results and compared to other approaches, in this section.
\subsection{Simulation Setup}
To simulate and assess the performance of our proposed approach, we used FLoRa for our simulations~\cite{adr-avg}. FLoRa (Framework for LoRa) is a simulation framework used for end-to-end simulations of LoRa networks. It uses OMNeT++, a discrete event network simulator, and is built on the INET framework. We customized FLoRa to simulate the resource allocation problem based on a novel low-complexity approach with low space complexity\footnote{The framework is available at
\url{https://github.com/reza-serati/ADR-Lite}.}.
\par Our study considers a LoRa network environment comprising a square-shaped cell with a dimension of 9800 m, which contains one GW in its center. As in~\cite{adr-avg}, in each simulation, the number of uniformly distributed EDs can vary from 100 to 700. We run the simulations for $12$ days while the channel saturation, i.e., sigma ($\sigma$), is equal to 7.08, and the $Oulu$ $Lora$ $Path$ $Loss$ model~\cite{pathloss} is considered for path loss. 
An interval time based on an exponential distribution with a mean of 1000 seconds is used between each transmission of a 20-Byte LoRa packet by each ED.
We average over 25 rounds within each scenario with randomly generated EDs' locations to obtain our simulation results.
\par To evaluate the proposed algorithm performance and compare
it with other schemes, we use the following metrics:
\begin{enumerate}
    \item \textbf{PDR}: This shows how many packets the GW has received, divided by how many packets were sent from each ED.
    \item \textbf{EC}: Calculated by dividing the total energy consumption by the PDR.
\end{enumerate}
In this study, we use the ratio of the total consumed energy by all EDs (which is calculated according to (\ref{e-toa})) to the PDR, as the performance measure for comparison purpose, that is:
\begin{equation} 
    EC = \frac{total~consumed~energy~by~all~EDs}{PDR}\,.
\label{per_pdr}
\end{equation}{}
\par We also consider the following four scenarios:
\begin{itemize}
    \item \textit{\textbf{Scenario 1:}} The number of nodes varies between 100 and 700, while EDs remain static.
    \item \textit{\textbf{Scenario 2:}} In addition to the EDs being mobile, the number of nodes varies between 100 and 700.
    \item \textit{\textbf{Scenario 3:}} The channel saturation value for 100\break static EDs takes the following values: $\{0, 0.89, 1.78, 2.67, 3.56, 4.46, 5.36, 6.24, 7.08\}$.
    \item \textit{\textbf{Scenario 4:}} In contrast to all other ADR algorithms, ADR-Lite allows not only SF and TP to be changed using the values: $\{7,8,9,10,11,12\}$ and $\{2,5,8,11,14\}$, respectively, but also CR and CF using the values: $\{\frac{4}{5}, \frac{4}{6}, \frac{4}{7}, \frac{4}{8}\}$ and $\{868.1, 868.4, 868.7\}$, respectively.
\end{itemize} 
\begin{figure}[t]
    \centering
    \includegraphics[width=\linewidth]{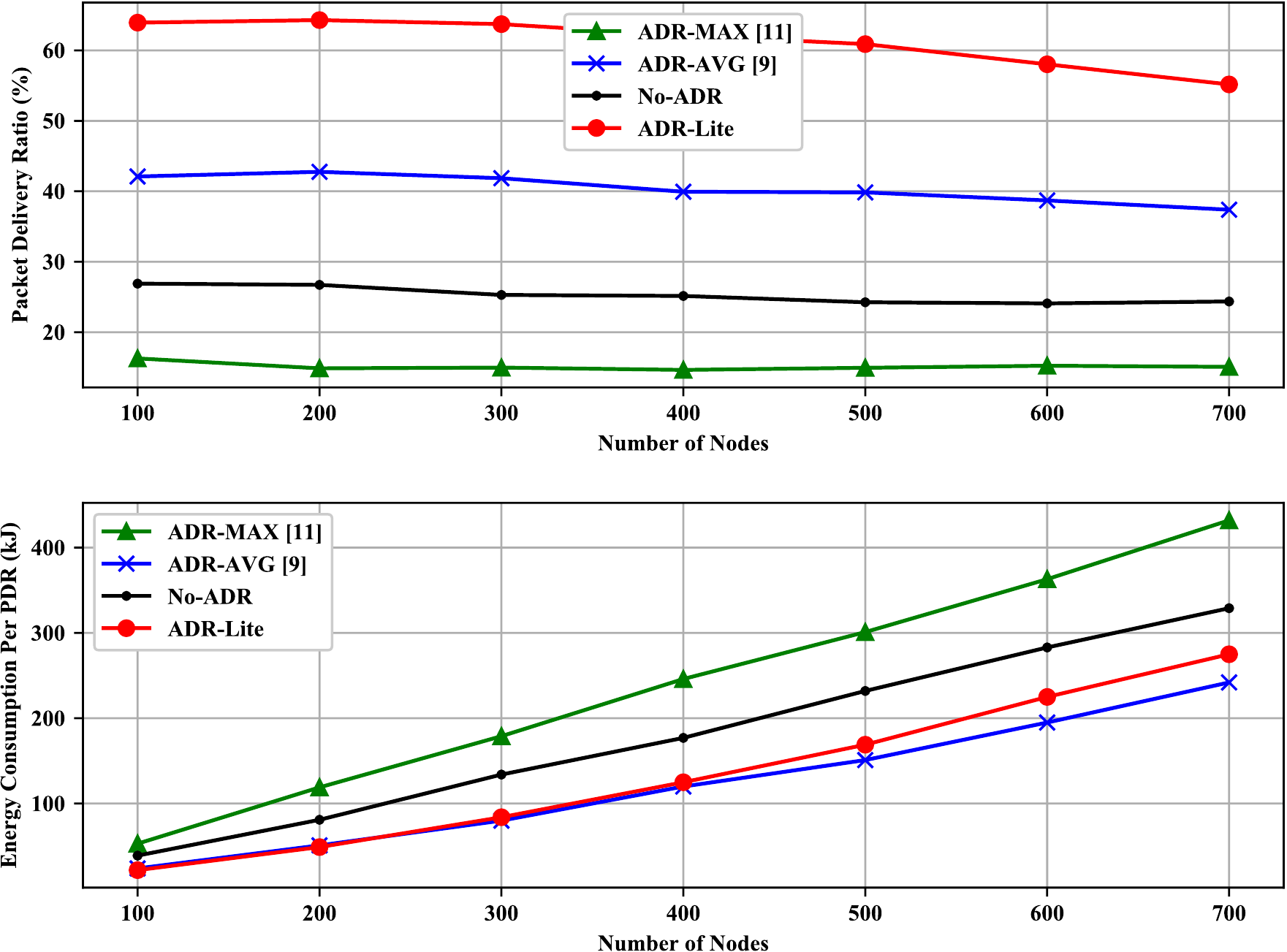}
    \caption{PDR \& EC of different algorithms versus number of static EDs in scenario 1 ($\sigma=7.08$).}
    \label{s1}
\end{figure} 
\subsection{Simulation Results}
Our simulation results compare the performance of our proposed algorithm with the ADR-MAX, ADR-AVG, and No-ADR schemes in the aforementioned four scenarios. 
\subsubsection{Scenario 1}
In this scenario, we assess the node number effect while every ED is assumed to be static. Figure \ref{s1} shows the PDR and energy consumption in the ADR-Lite, ADR-MAX, ADR-AVG, and No-ADR algorithms in scenario 1. Due to the fact that ADR-Lite only stores the details of the last received packet, the current status of the link is realistically considered. This greedy manner of decision-making leads to focusing on the packet reception as the primary objective. Thus, as seen, a higher level of PDR regardless of network densification can be achieved.
\par Despite the fact that our proposed solution does not avoid using high-TP transmission configurations, aiming for more packet reception, ADR-Lite does not have the minimum energy consumption in a dense network deployment, whereas it does provide the least energy consumption in a less crowded environment.
\begin{figure}[t]
    \centering
    \includegraphics[width=\linewidth]{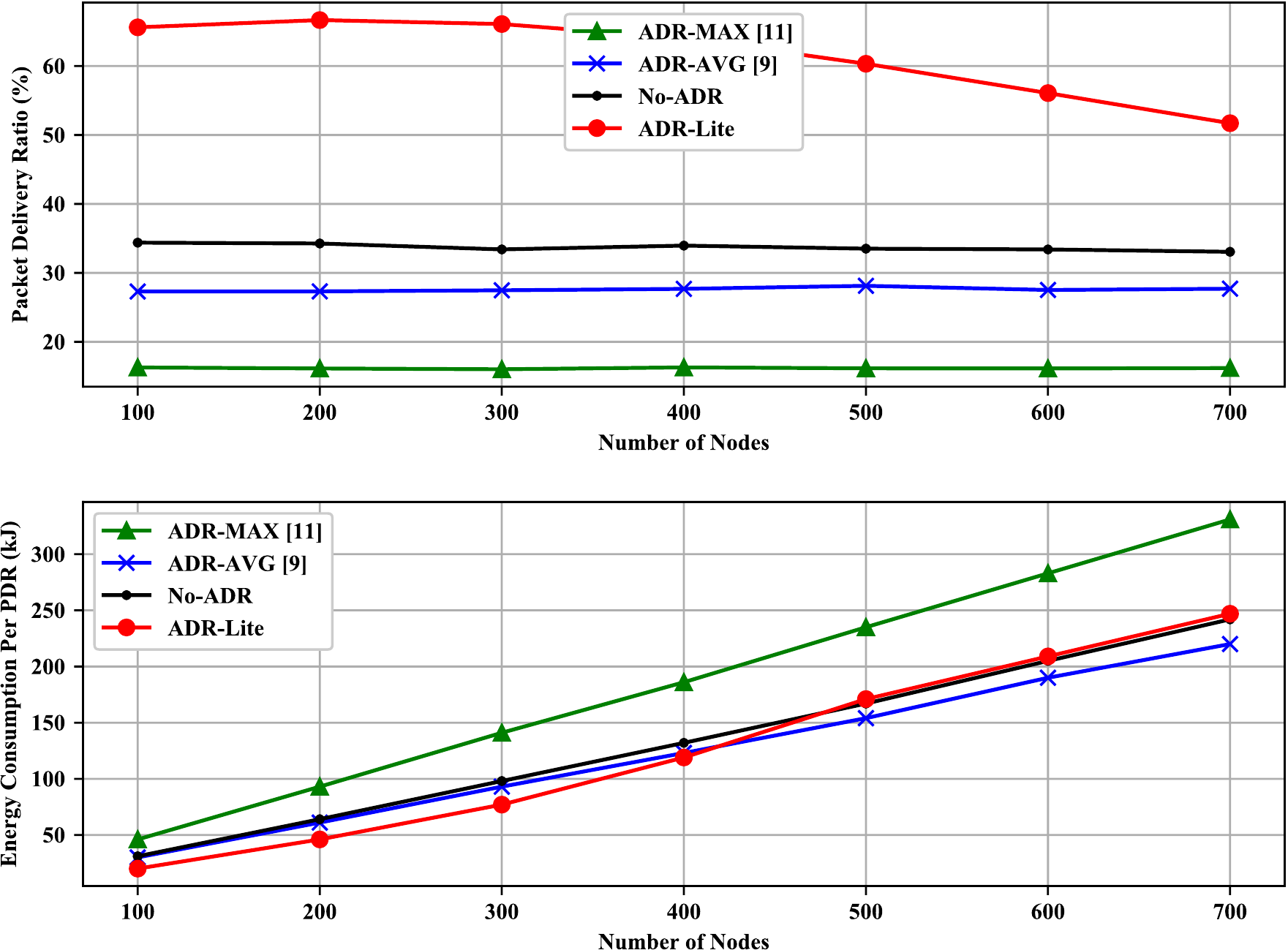}
    \caption{PDR \& EC of different algorithms versus number of mobile EDs in scenario 2.}
    \label{s2}
\end{figure} 
\begin{figure}[b]
    \centering
    \includegraphics[width=\linewidth]{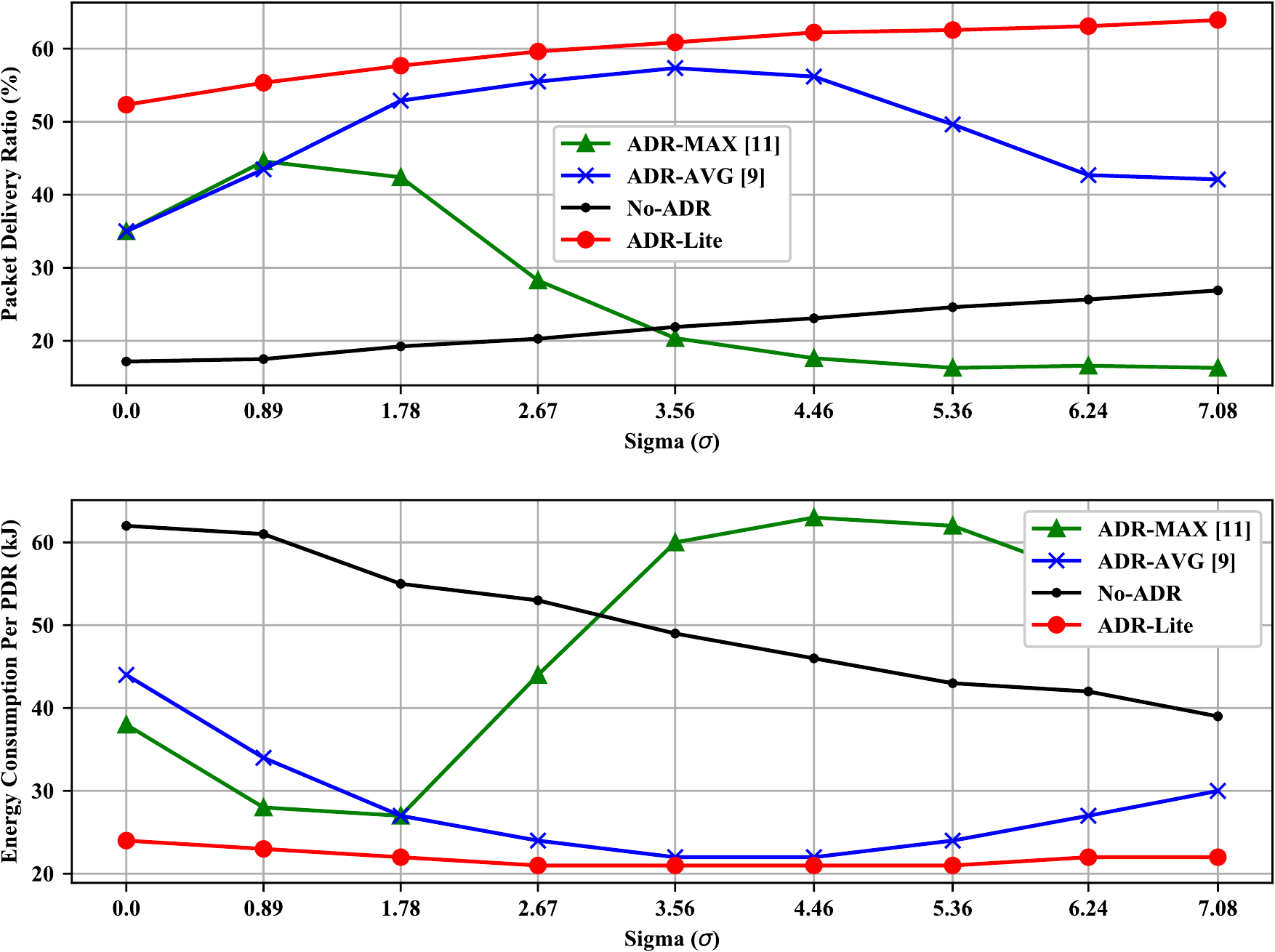}
    \caption{PDR \& EC of different algorithms versus $\sigma$ in scenario 3 (100 static EDs).}
    \label{s3}
\end{figure}
\subsubsection{Scenario 2} 
The IoT has enabled a wide variety of applications that demand or apply mobility. We see examples of mobile use cases in smart vehicles, smart cities, and smart health-care applications~\cite{iot}. Thus, this scenario aims to investigate the effects of mobility on the performance of EDs while utilizing various ADR mechanisms. In scenario 2, Figure \ref{s2} illustrates the PDR and energy consumption of mobile EDs for the ADR-Lite, ADR-MAX, ADR-AVG, and No-ADR algorithms. This work used the random waypoint mobile model, which has been widely used to simulate ad hoc networks, in order to represent the motion of EDs having a speed between 0 and 5~m/s that follows an exponential distribution.
\par Despite maintaining the highest PDR in any ED densification, ADR-Lite experiences a downward trend as more EDs are added to the network, followed by an increase in energy consumption. The reason is that, as this method does not avoid high TP selections, aiming to increase packet reception, EDs further away from the GW are unable to send packets to the GW successfully when EDs closer to it are transmitting with their full TP, resulting in destructive collisions. This phenomenon is also known as the near-far problem, which may reduce PDR and increase energy consumption.
\subsubsection{Scenario 3} 
The PDR and energy consumption of different algorithms versus $\sigma$ for the 100 static EDs used in scenario 3, is illustrated in Figure \ref{s3}. Due to the fact that ADR-Lite transmits parameters without considering the history of recent packets, it achieves a higher PDR than any other scheme, which even increases slightly as channel saturation $\sigma$ increases. The energy consumption of ADR-Lite is almost steady as $\sigma$ varies. On the contrary, with $\sigma$ increasing, the ADR-MAX's energy consumption largely increases in the noisy channel. The reason is that, as $\sigma$ increases, the ADR-MAX's PDR decreases, which causes the EDs to transmit with more power, leading to an increasing energy consumption.
\subsubsection{Scenario 4} 
In comparison to the other approaches, which only configure SF and TP to find an optimal network performance, ADR-Lite allows EDs to also accept different values for other transmission parameters such as CR and CF. This can increase freedom of choice by enhancing the state space. In addition, the proposed approach does not add overhead to the LoRa header packet and makes no changes to the protocol design. Unlike previous scenarios, the simulation takes place over 120 days, while the ADR-Lite's combinations of parameters varies over $\{30, 90, 120, 360\}$ selections. For this scenario, we assume four different configurations, namely config-1, config-2, config-3, and config-4, where the transmission parameters are: $\{SF+TP\}$, $\{SF+TP+CF\}$, $\{SF+TP+CR\}$, and $\{SF+TP+CF+CR\}$, respectively. However, it is important to note that the CF may not be adjustable for each ED in real environments, but it can be modified using the FLoRa simulation framework.
\par
\begin{figure}[t]
    \centering
    \includegraphics[width=\linewidth]{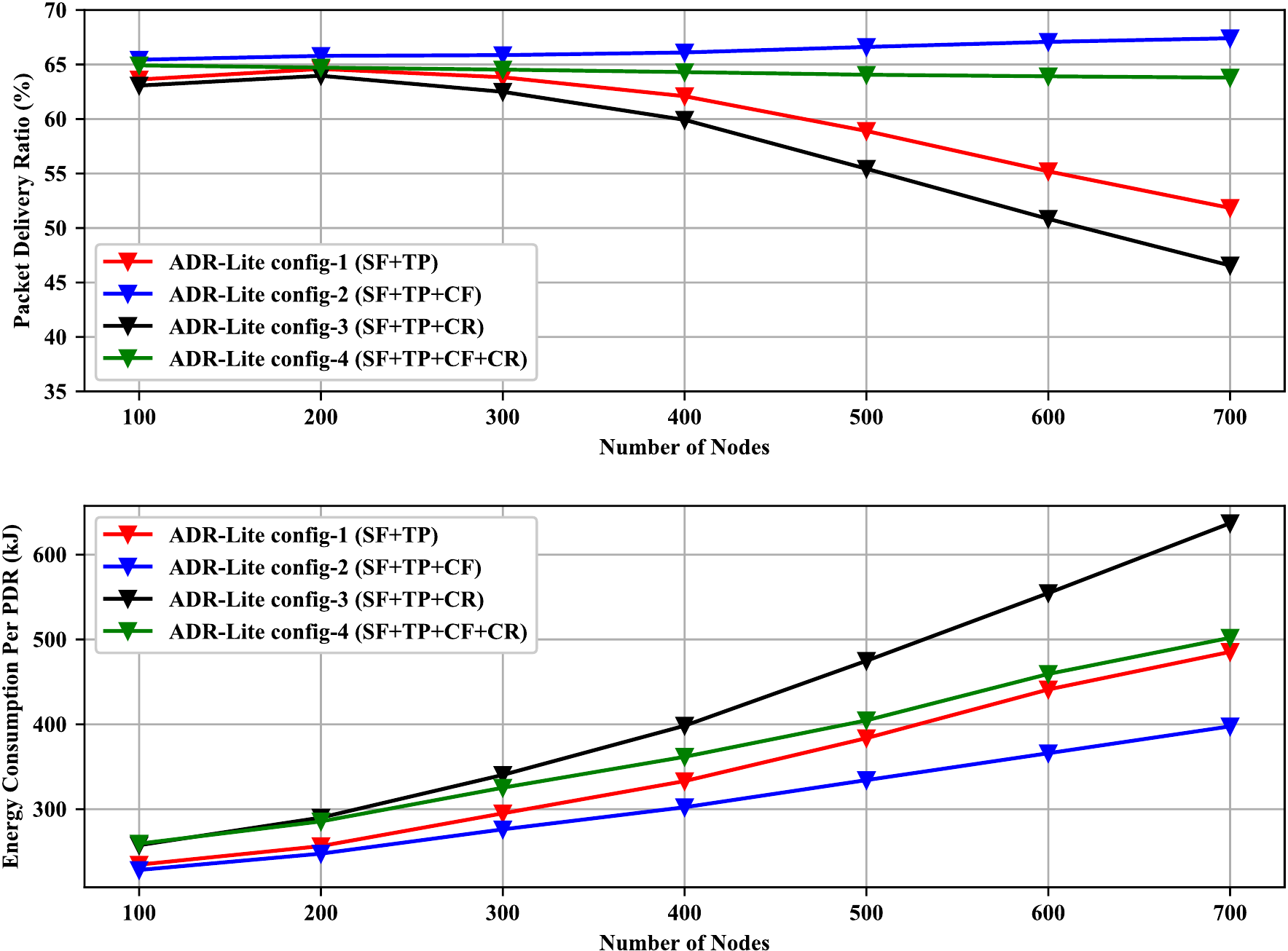}
    \caption{PDR \& EC of different configuration versus number of static EDs in scenario 4.}
    \label{s4}
\end{figure}

 In  Figure \ref{s4}, we can see that a greater degree of freedom of choice generally leads to better network performance, both in terms of PDR and energy consumption. As expected, the PDR and energy consumption results for config-1, which comprises the default configuration parameters, are the same as for scenario 1. Due to the fact that config-2 adds more channels to the communication link while adding no overhead to the network, it can result in the best performance for both PDR and energy consumption compared to the other configurations. In config-3, since the default CR is~$\frac{4}{5}$, by increasing this parameter's selection state space, the total overhead of the network will grow, which can result in longer transmission delays, thus increasing the collision rate and energy consumption. Config-4's performance is slightly disappointing compared to config-2, because although more channels can improve efficiency, a wide range of error coding options can negatively impact performance.
\section{Conclusion and Future Works}
In this paper, we proposed a low-complexity ADR algorithm for the NS of LoRa networks that can allow static or mobile EDs to configure the transmission parameters of SF, TP, CF, and CR. ADR-Lite uses a binary search algorithm to find the optimal configuration of transmission parameters for a LoRa network after sorting the parameter values in ascending order. The proposed algorithm, without considering the history of the last received packets, attempts to obtain the optimal transmission parameter values. Through simulation results, we showed the effectiveness of the proposed algorithm in a LoRa network which outperformed others in static or mobile scenarios with different network densification and channel conditions. As future work, the packet retransmission method can be included to increase the PDR further while employing machine learning approaches to find an optimal EC without any PDR reduction.
\balance


\end{document}